\documentclass[final]{eps}
\usepackage{graphicx}
\usepackage{times}
\newcommand       \Angstrom     {\,{\rm \AA}}
\newcommand       \um           {\mu{\rm m}}
\newcommand       \mum          {\,{\rm \mu m}}
\newcommand       \g            {\,{\rm g}}
\newcommand       \cm           {\,{\rm cm}}
\newcommand       \tausil       {\Delta\tau_{9.7\mum}}
\newcommand       \tauahc       {\Delta\tau_{3.4\mum}}
\newcommand       \simali       {{\sim}\,}
\newcommand       \magni        {\,{\rm mag}}
\newcommand       \sgrA         {{\rm Sgr}\,{\rm A}^{\ast}}
\volumenumber{58} \publishyear{2008} \frompage{\pageref{firstpage}}
\topage{\pageref{finalpage}}

\shorttitle{Gao, Jiang, \& Li: On the 3.4$\mum$ and 9.7$\mum$
Interstellar Extinction Variations}

\title{Toward understanding the 3.4$\mum$ and 9.7$\mum$ extinction
feature variations from the local diffuse interstellar medium to
the Galactic center}

\author{Jian~Gao$^{1,2}$, B.~W.~Jiang$^1$, and Aigen~Li$^2$}
\affiliation{$^1$Department of Astronomy,
                 Beijing Normal University,
                 Beijing, 100875, China\\
             $^2$Department of Physics and Astronomy,
                 University of Missouri,
                 Columbia, Missouri 65211, USA\\}
\received{August 10, 2008}\revised{November 28,
2008}\accepted{December 4, 2008}

\abstract{%
Observationally, both the 3.4$\mum$ aliphatic hydrocarbon C--H
stretching absorption feature and the 9.7$\mum$ amorphous silicate
Si--O stretching absorption feature show considerable variations
from the local diffuse interstellar medium (ISM) to Galactic center
(GC): both the ratio of the visual extinction ($A_V$) to the
9.7$\mum$ Si--O optical depth ($\tausil$) and the ratio of $A_V$ to
the 3.4$\mum$ C--H optical depth ($\tauahc$) of the solar
neighborhood local diffuse ISM are about twice as much as that of
the GC. In this work, we try to explain these variations in terms of
a porous dust model consisting of a mixture of amorphous silicate,
carbonaceous organic refractory dust (as well as water ice for the
GC dust).
}


\begin{document}
\label{firstpage}
\maketitle
\copyrighttext{}

\section{Introduction}
The interstellar extinction law is one of the primary sources
of information about the interstellar grain population, and one
often obtains direct information on the composition of interstellar
dust from spectral features in extinction (Draine 2003). These
spectral features also provide strong constraints on interstellar
grain models. With the advent of ground-based and space borne
infrared (IR) telescope facilities, the IR extinction continuum
and absorption features have been receiving increasing attention
and play an essential role in recovering the intrinsic energy
distribution of celestial objects
and inferring the characteristics of interstellar dust.

In the interstellar extinction curve, the 2175$\Angstrom$
bump is outstanding in the ultraviolet (UV),
while in the IR, there are a number of prominent
absorption features as well:
(1) the ubiquitous 9.7$\mum$ and 18$\mum$ features respectively
    due to the Si--O stretching and O--Si--O bending modes
    of amorphous silicates;
(2) the 3.4$\mum$ feature due to the C--H stretching mode
    of aliphatic hydrocarbon dust, as ubiquitously present in
    the ISM of the Milky Way and external galaxies
    as the 9.7$\mum$ and 18$\mum$ silicate bands,
    except this feature is not seen in dense molecular clouds
    (see Pendleton 2004 for a review);
(3) the 3.3$\mum$ and 6.2$\mum$ weak features
    seen in both local sources and GC sources
    (Schutte et al.\ 1998; Chiar et al.\ 2000),
    respectively due to the C--H stretching and C--C stretching
    modes of polycyclic aromatic hydrocarbon (PAH) molecules; and
(4) in dense clouds the 3.1$\mum$ feature due to the O--H stretching
    mode of water ice as well as a number of weaker features
    at 4.68$\mum$ (CO), 7.68$\mum$ (CH$_4$),
    4.28$\mum$, 15.2$\mum$ (CO$_2$),
    3.54$\mum$, 9.75$\mum$ (CH$_3$OH).

\begin{table*}[t] 
\renewcommand{\arraystretch}{1.2}
\vspace{0cm} \caption{Observational values of $A_V/\tauahc$ and
$A_V/\tausil$} \vspace{-0.1cm}
\begin{center}
\begin{tabular}{cccc}\hline\hline
& $A_V/\tauahc$ & $A_V/\tausil$ &  References\\
\hline & 240$\pm$40 & &  Sandford et al.\ 1991 \\
& 250$\pm$40 & &  Pendleton et al.\ 1994 \\
& 333 & 16.7   &  Adamson et al.\ 1990 \\
\raisebox{1.5ex}[0cm][0cm]{\textbf{Local diffuse ISM}} &  & 18.5  & Roche \& Aitken 1984 \\
&  & 18.5$\pm$2 & Draine 2003 \\
&  & 19.2$\pm$0.6& Bowey et al.\ 2004 \\
mean for local ISM &  274  & 18.2 & \\
\hline  &   150$\pm$20 &      &  Pendleton et al.\ 1994 \\
         & 143 &  8.31    & McFadzean et al.\ 1989 \\
\raisebox{1.5ex}[0cm][0cm] {\textbf{Galactic Center}}
&  & 8$\pm$3 & Becklin et al.\ 1978 \\
&  &  9 & Roche \& Aitken 1985 \\
mean for GC  & 146 & 8.4 &  \\
\hline
\end{tabular} \label{tab1}
\end{center}
\end{table*}

The 9.7$\mum$ silicate extinction profile varies among different
sightlines; in particular, its optical depth $\tausil$ (relative
to the visual extinction $A_V$) shows considerable variations
from the local diffuse ISM (LDISM) to the Galactic center (GC):
$A_V/\tausil \approx 18.2$ for LDISM differs from
that of the GC ($A_V/\tausil \approx 8.4$) by a factor
of $\simali$2.2 (see Table 1; also see Draine 2003).\footnote{%
  $^1$Cohen et al.\ (1989) argued that the observed
  9.7$\mum$ dip on which $\tausil$ was measured
  may be partly contributed by PAHs
  (i.e. the red tails of the PAH 7.7$\mum$ and 8.6$\mum$
   bands and the blue tail of the 11.3$\mum$ band could
   form an ``artificial'' 10$\mum$ dip).
  But this would result in a smaller $\tausil$
  for the GC and a larger $\tausil$ for the LDISM
  (since the PAH emission is more likely to present
   in the LDISM while toward the GC PAHs are seen
   in absorption), quite on the opposite.
  }
Roche \& Aitken (1985) argued that $A_V/\tausil$ varies
because there are fewer carbon stars in the central regions
of the Galaxy and the production of carbon-rich dust may be
substantially reduced compared with the outer Galactic disk.
However, as shown in Table~1, the 3.4$\mum$ C--H feature of
aliphatic hydrocarbon dust also exhibits a similar behavior:
$A_V/\tauahc \approx 274$ for LDISM is higher than
that of the GC ($A_V/\tauahc \approx 146$) by a factor
of $\simali$1.9 (see Table 1).
If the argument of Roche \& Aitken (1985) was valid,
one would expect a much smaller $A_V/\tauahc$ ratio
in the LDISM than that of the GC.
Sandford et al.\ (1995) tried to quantitatively
explain this phenomena by assuming that the abundance
of the C--H carrier (relative to other dust components)
gradually increases from the local ISM toward the GC.\footnote{%
  $^2$They also pointed out that the C--H and Si--O carriers
  may be coupled, perhaps in the form of silicate-core
  organic-mantle grains.
  This idea is challenged by the nondetection of
  the 3.4$\mum$ feature polarization along sightlines
  where the 9.7$\mum$ feature polarization is detected
  (Adamson et al.\ 1999, Chiar et al.\ 2006;
  also see Li \& Greenberg 2002).
  It is possible that the 3.4$\mum$ feature may be not produced
  by a carrier residing in a mantle on a silicate core
  but by very small (unaligned) grains (Chiar et al.\ 2006).
  }
However, this requires that amorphous silicate dust
and aliphatic hydrocarbon dust should not be solely
responsible for the visual extinction.
If one has to invoke an additional dust component
(most likely a population of carbon dust which does
not show the characteristic 3.4$\mum$ feature, say,
graphite) making an appreciable contribution to $A_V$,
one would encounter a severe carbon budget problem
(see Snow \& Witt 1996).

Along the lines of sight toward the GC,
there are dense molecular clouds.\footnote{%
  $^3$The sightline toward the Galactic center source
  $\sgrA$ suffers about $\simali$30$\magni$ of
  visual extinction (e.g. see McFadzean et al.\ 1989),
  to which molecular clouds may contribute as much as
  $\simali$10$\magni$ (Whittet et al.\ 1997).
  }
In cold, dense molecular clouds, interstellar dust is expected to
grow through coagulation (as well as accreting an ice mantle) and
the dust is likely to be porous (Jura 1980). In this work, we
demonstrate that the observed variations of $A_V/\tausil$ and
$A_V/\tauahc$ from the LDISM to the GC could be explained in terms
of composite porous dust.

\section{Model}
We consider a composite porous dust model consisting of amorphous
silicate, carbon, and vacuum (in dense clouds silicate dust and
carbon dust are coated with water ice). We take the optical
constants of Draine \& Lee (1984) for amorphous silicate, of Li \&
Greenberg (1997) for carbonaceous organic refractory (to represent
the carbon dust component), of Li \& Greenberg (1998) for water ice.
The mass densities of silicate dust, organic refractory dust and ice
are taken to be $\rho_{\rm sil}\approx3.5\g\cm^{-3}$, $\rho_{\rm
carb}\approx1.8\g\cm^{-3}$ and $\rho_{\rm ice}\approx1.2\g\cm^{-3}$,
respectively. We take the mass ratio of organic refractory dust to
silicate dust to be $m_{\rm carb}/m_{\rm sil} = 0.7$ and the mass
ratio of water ice to organic refractory dust and silicate dust to
be $m_{\rm ice}/\left(m_{\rm carb}+m_{\rm sil}\right) = 0.8$, as
inferred from the cosmic abundance constraints (see Appendix A of Li
\& Lunine 2003a).
\begin{figure}[t]
\vspace{-0.7cm} \centerline{\vspace{-2.5cm}
\includegraphics[scale=.38,clip]{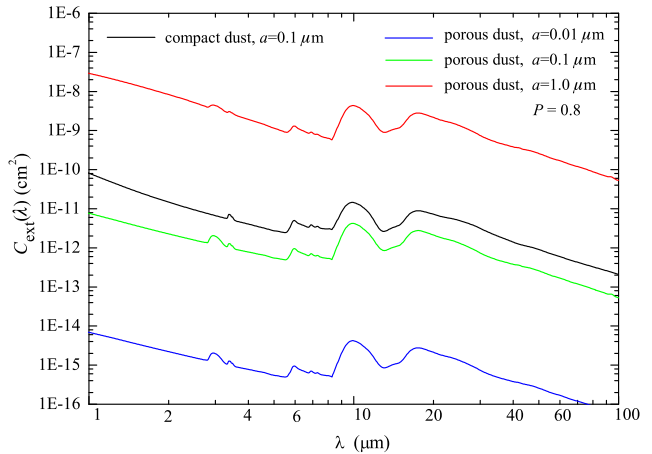}}
\vspace{1.8cm}
\caption{%
         Extinction cross sections $C_{\rm ext}(\lambda)$
         of different types of dust.
         The 3.1$\mum$ water ice O--H feature
         shows up in the extinction profiles of porous dust.
         \label{fig1}}
\end{figure}

\begin{figure*}[t]
\centerline{\includegraphics[angle=90,scale=.33]{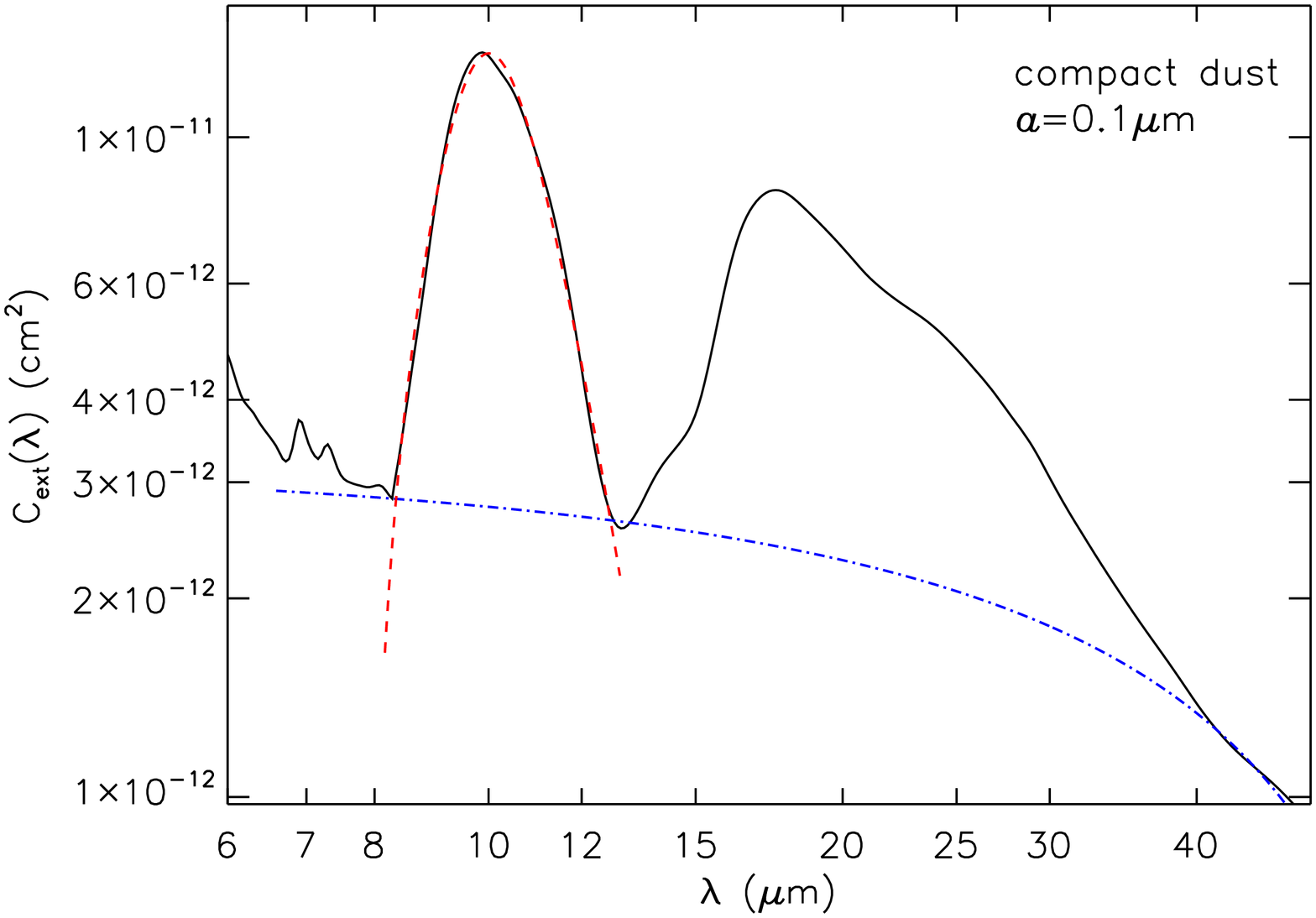}\hspace{0.8cm}
\includegraphics[angle=90,scale=.33]{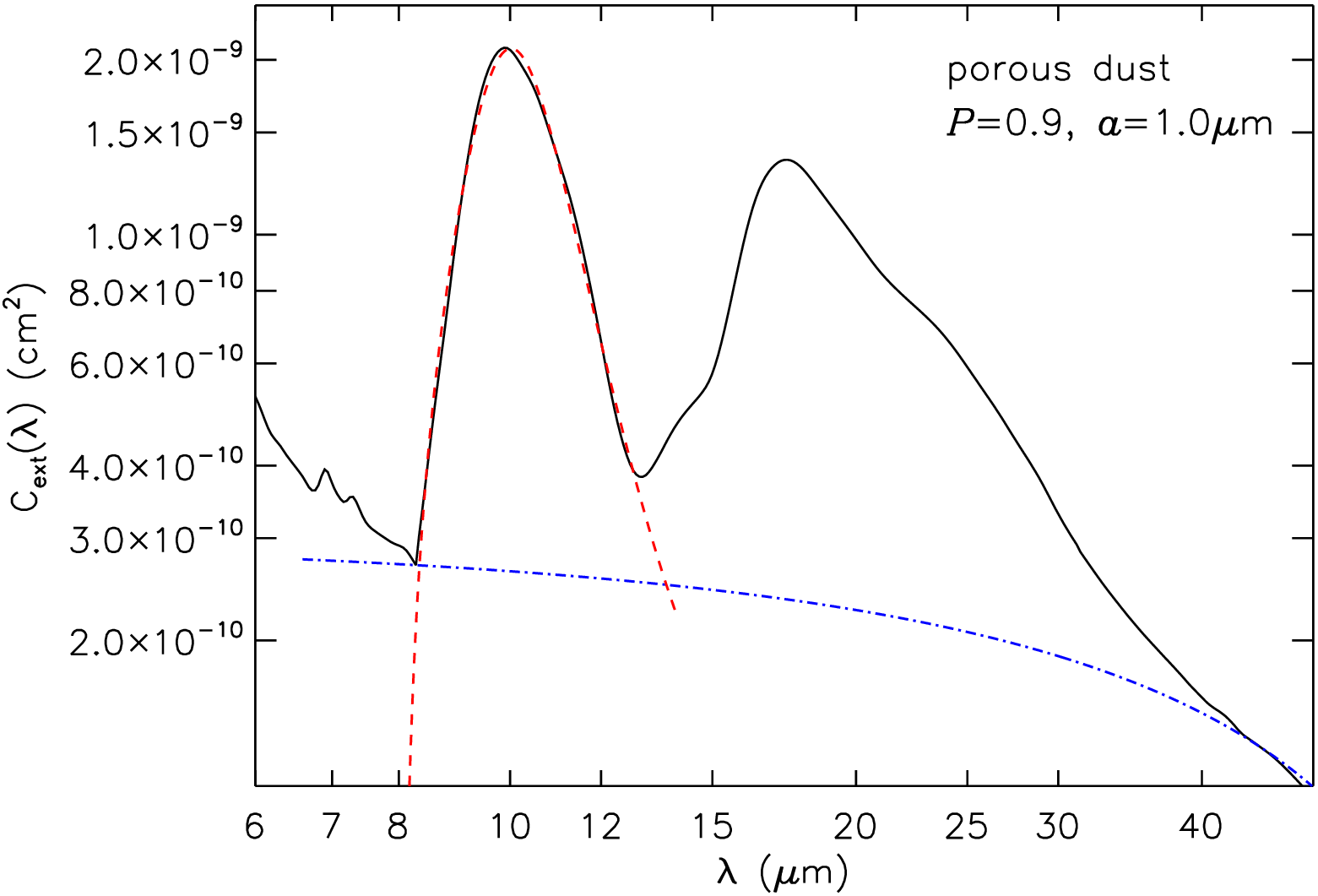}}
\caption{%
         Schematic illustration of obtaining
         $\Delta C_{\rm ext}(9.7\mum)$,
         the excess 9.7$\mum$ extinction cross section
         above the continuum. It is obtained
         by (1) fitting the 9.7$\mum$ model profile
         with a Drude function and subtracting
         the continuum which is fitted with
         a six-order polynomial, (2) integrating
         the continuum-subtracted extinction profile
         (which is approximated by a Drude function)
         over wavelength, and finally (3) dividing
         the integrated value with the width of
         the interstellar 9.7$\mum$ silicate absorption feature.
         \label{fig2}}
\end{figure*}

\begin{figure*}[t]
\centerline{\includegraphics[angle=90,scale=.33]{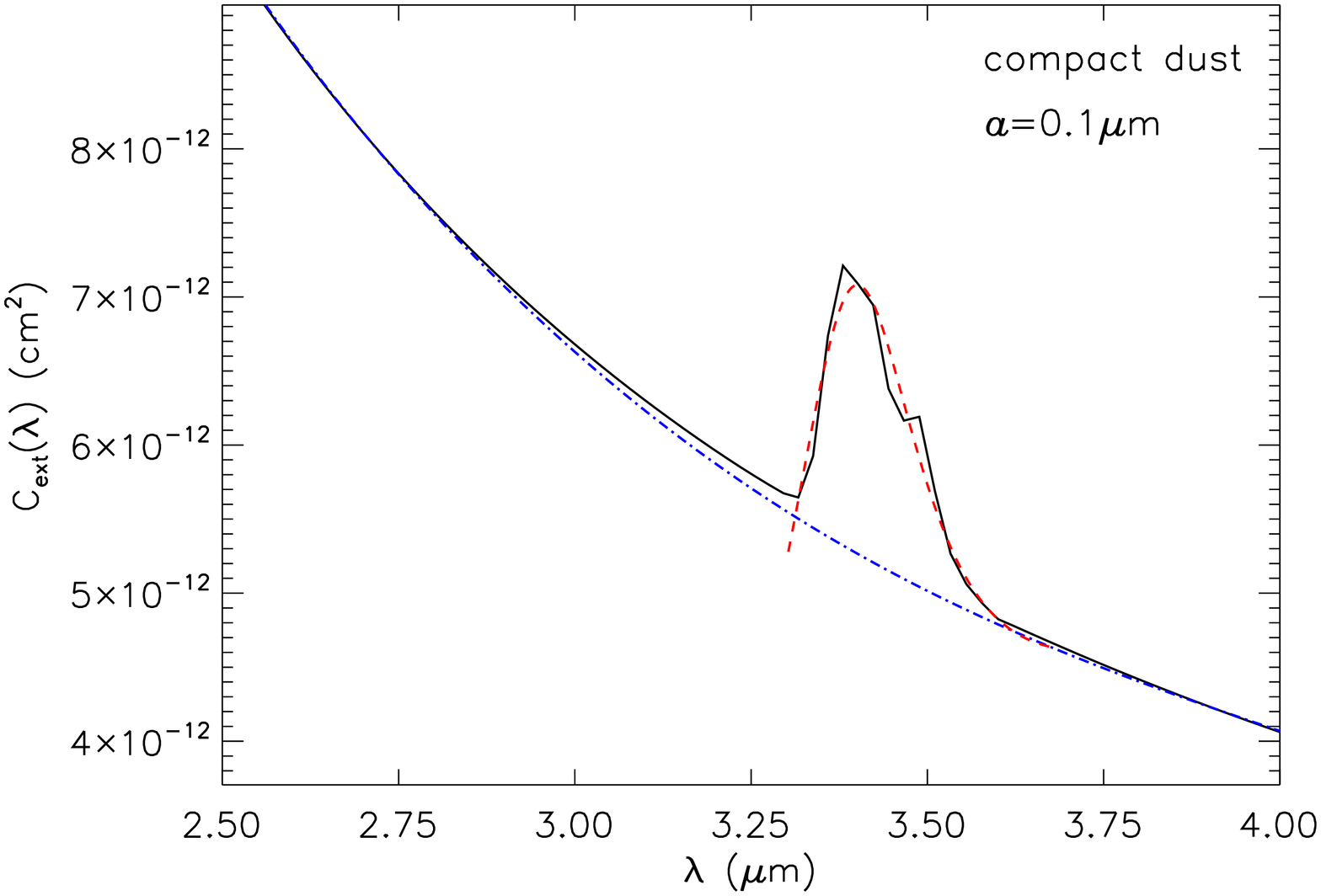}\hspace{0.8cm}
\includegraphics[angle=90,scale=.33]{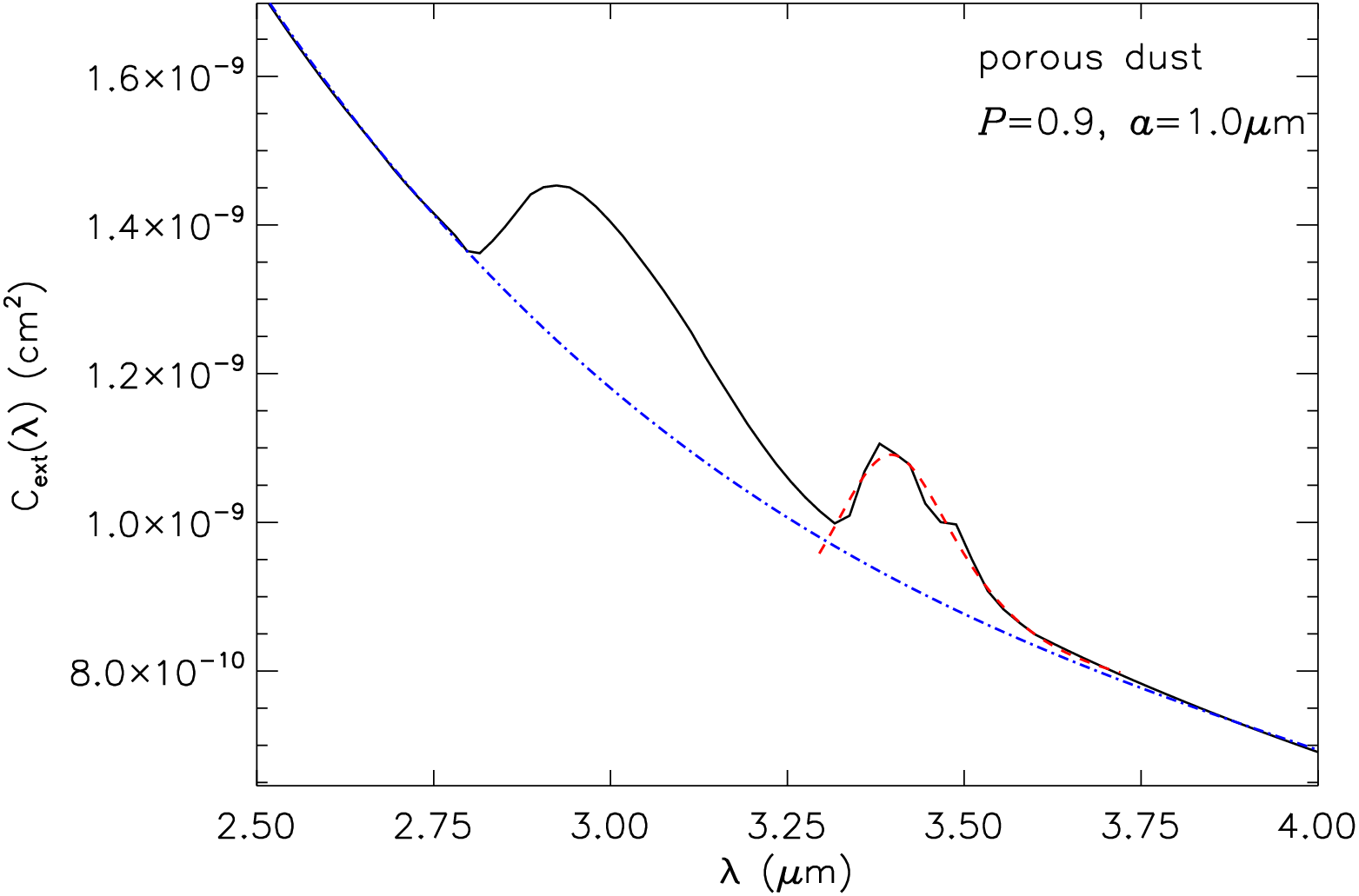}}
\caption{%
         Same as Figure~2 but for the 3.4$\mum$ feature.
         In the right panel, the 3.1$\mum$ water ice O--H feature
         shows up in the extinction profile of porous dust.
         \label{fig3}}
\end{figure*}

For the dust in the local diffuse ISM, we assume the dust to be a
solid compact mixture of amorphous silicate and organic refractory
materials with $m_{\rm carb}/m_{\rm sil} = 0.7$. We take the dust
size to be $a=0.1\mum$, the typical grain size for the dust in the
diffuse ISM (see Draine 1995). For the dust in the dense molecular
clouds along the lines of sight toward the GC, we assume that
silicate dust and organic refractory dust are equally coated with an
ice layer and then form a porous aggregate (see Li \& Lunine 2003b).
For porous dust, a key parameter is the porosity $P$ (or fluffiness;
the fractional volume of vacuum in a grain). We will consider a
range of porosities. We assume all grains are spherical in shape;
the porous grain size $a$ is defined as the radius of the sphere
encompassing the entire porous aggregate. In order to find suitable
porosity $P$ and dust size $a$ for the dust in the dense molecular
clouds to reproduce the observed $A_V/\tausil$ and $A_V/\tauahc$
ratios toward the GC, we leave both $P$ and $a$ adjustable.

We use Mie theory in combination with the Maxwell-Garnett and
Bruggeman effective medium theories (Bohren \& Huffman 1983; see
eqs.7--9 of Li \& Lunine [2003b] and Kimura et al.\ 2008b) to
calculate the optical properties of composite porous grains. This
approach is valid for computing the integral scattering
characteristics (e.g. extinction, scattering, absorption cross
sections, albedo and asymmetry parameter; see Hage \& Greenberg
1990, Wolff et al.\ 1994).

For illustration, we plot in Figure~1 the 1--100$\mum$ extinction
cross sections $C_{\rm ext}(\lambda)$ of compact dust (for the local
diffuse ISM) and porous dust (for dense clouds toward the GC). For a
given dust size, both $A_V$ and $\tauahc$, $\tausil$ decrease with
the porosity $P$, although the degree of decrease is somewhat less
significant for $\tauahc$ and $\tausil$ than for $A_V$ (thus
$A_V/\tauahc$ and $A_V/\tausil$ decrease moderately with the
increase of $P$; see Table~2). This is because with the same size, a
porous grain contains less dust material than its solid counterpart.
The effective dielectric functions of porous dust are reduced. In
the IR the extinction for dust of $\sim$\,0.1$\mum$ is dominated by
absorption (i.e. in the Rayleigh regime) and is roughly proportional
to the imaginary parts of the dielectric functions (see Li 2008).
Therefore porous dust produces smaller $\tauahc$ and $\tausil$ than
its solid counterpart of the same size. In the optical, both
absorption and scattering are important. The introduction of vaccum
leads to a reduction of the dielectric functions of the dust which
will decrease both the absorption and scattering.

\begin{table*}[t] 
\renewcommand{\arraystretch}{1.2}
\vspace{-.3cm}
\caption{%
         $A_V/\tauahc$ and $A_V/\tausil$ calculated
         for various dust models
         }
\vspace{-.1cm}
\begin{center}
\begin{tabular}{ccccccc}
\hline \hline Dust  & Porosity  & size & $A_V/\tauahc$
     & FWHM & $A_V/\tausil$ & FWHM  \\
       Model  &  $P$ & ($\um$) &  & ($\um$)  & & ($\um$) \\
\hline diffuse ISM Dust\tablenotemark{a}
& 0& 0.1 & 243& 0.14 & 38.2&2.18 \\
& 0.1& 0.1 & 216 & 0.14 & 32.3 &2.19 \\
& 0.2& 0.1 & 192 & 0.14 & 27.3 &2.19 \\\hline
 & & 0.1 & 80 & 0.14& 5.9 & 2.18\\
 & &0.5& 150 & 0.15 & 13.3 &2.18 \\
 & \raisebox{-2ex}[0cm][0cm]{0.8}&1.0 & 186 & 0.17 & 16.3 & 2.18 \\
 & & 1.5 & 127 & 0.19 & 14.8 & 2.21 \\
 & & 2.0& 102 & 0.19& 11.5 & 2.26 \\
\raisebox{-2ex}[0cm][0cm]{GC Dust} & & 3.0& 55 & 0.20& 6.1& 2.34 \\
  \cline{2-7} & & 0.1& 72 & 0.14& 5.1 &2.08 \\
 & & 0.5& 126& 0.14 & 9.3&2.08 \\
 & \raisebox{-2ex}[0cm][0cm]{0.9} & 1.0& 165& 0.15 & 12.8 &2.09 \\
 & & 1.5& 193& 0.16 & 15 &2.11 \\
 & & 2.0& 198 & 0.15 & 15.6 &2.11\\
 & & 3.0& 174 & 0.16 & 14.1 &2.16 \\
\hline
\end{tabular}\label{tab2}
\end{center}
\tablenotemark{a}{Diffuse ISM dust is taken to be a mixture of
                  amorphous silicate and organic refractory substance.}
\end{table*}

For models consisting of single-sized dust, the ratio of $A_V$ to
the optical depth of the 9.7$\mum$ silicate feature is simply
$A_V/\tausil \approx 1.086\,C_{\rm ext}(V)/\Delta C_{\rm
ext}(9.7\mum)$, where $C_{\rm ext}(V)$ is the extinction cross
section at $V$ band ($\lambda = 5500\Angstrom$), $\Delta C_{\rm ext
}(9.7\mum)$ is the {\it excess} extinction cross section of the
9.7$\mum$ feature above the continuum, and the factor ''1.086''
arises from the conversion of extinction (in magnitude) to optical
depth. We obtain $\Delta C_{\rm ext}(9.7\mum)$ by integrating the
9.7$\mum$ model extinction profile (with the continuum subtracted)
over wavelength (see Fig.\,2 for illustration) and then dividing the
integrated value with the width of the interstellar 9.7$\mum$
silicate absorption feature. The same procedure is applied to the
3.4$\mum$ feature to calculate $\Delta C_{\rm ext}(3.4\mum)$ so as
to obtain $A_V/\tauahc$ (see Fig.\,3 for illustration).

\section{Results and Discussion}
We calculate $A_V/\tauahc$ and $A_V/\tausil$ for various dust models
with a range of porosities and dust sizes. The results are tabulated
in Table~2. For a given dust size $a=0.1\mum$, porous dust results
in smaller $A_V/\tauahc$ and $A_V/\tausil$ values than compact dust
(see \S2). This is encouraging that porous dust which is likely
present in the dense molecular clouds along the sightlines toward
the GC indeed is on the right track to account for the observed
$A_V/\tauahc$ and $A_V/\tausil$ variations from the local diffuse
ISM to the GC. More specifically, from Table~2 we see that the dust
with $P\sim 0.8-0.9$ and $a\sim 0.5-1\mum$ can approximately explain
the observed variations of $A_V/\tauahc$ and $A_V/\tausil$ (by a
factor of $\simali$2) from the local diffuse ISM to the GC. Both
high porosities ($P\sim 0.8-09$) and large sizes ($a\ge 0.5\mum$)
are required for the GC dust to account for the lower $A_V/\tauahc$
and $A_V/\tausil$ ratios.

In Figure~4, we show $A_V/\tauahc$ and $A_V/\tausil$ as a function
of dust size for $P=0.8,\,0.9,0.95$. It is clearly seen that for a
given porosity, the variation of $A_V/\tauahc$ with dust size
exhibits a tendency similar to that of $A_V/\tausil$. This suggests
that with a dust size distribution taken into account, we would
still maintain the variation tendency. For small, highly porous
grains ($a<0.05\mum$), $A_V/\tauahc$ and $A_V/\tausil$ are nearly
independent of size since they are more or less in the Rayleigh
regime in the optical-IR and therefore $A_V/V_{\rm dust}$,
$\tauahc/V_{\rm dust}$ and $\tausil/V_{\rm dust}$ are independent of
grain size (where $V_{\rm dust}$ is the dust volume). The visual
extinction $A_V$ reaches its maximum at $a\sim
\lambda/\left[2\pi\,a\,(n-1)\right]$ (where $n$ is the real part of
the refractive index of the porous dust at wavelength $\lambda$; see
Li 2008). At even large sizes, while $A_V$ reaches a constant (i.e.
in the geometric optics regime) $\tauahc$ and $\tausil$ increase
with $a$ (till they reach their respective peak values). This
explains why $A_V/\tauahc$ and $A_V/\tausil$ decrease with $a$ after
they reach their peak values.

For $A_V/\tauahc$, our model with $m_{\rm carb}/m_{\rm sil} = 0.7$
(and $P=0.8$, $a=0.5-1.5\mum$) is consistent with the observed
factor-of-two variations in the local ISM and toward the GC (see
Tables~1,2). However, the model values of $A_V/\tausil$ for both the
diffuse ISM dust ($A_V/\tausil\approx 38.2$) and the GC dust
($A_V/\tausil\approx 16.3$) are higher by a factor of
$\simali$1.5--2 than that observed in the local diffuse ISM
($A_V/\tausil\approx 18.2$) and the GC ($A_V/\tausil\approx 8.4$).
This discrepancy may result from the underestimation of the silicate
mass fraction.

\begin{figure}[t]
\vspace{-0.8cm}
\centerline{\hspace{0.2cm}\includegraphics[scale=.38,clip]{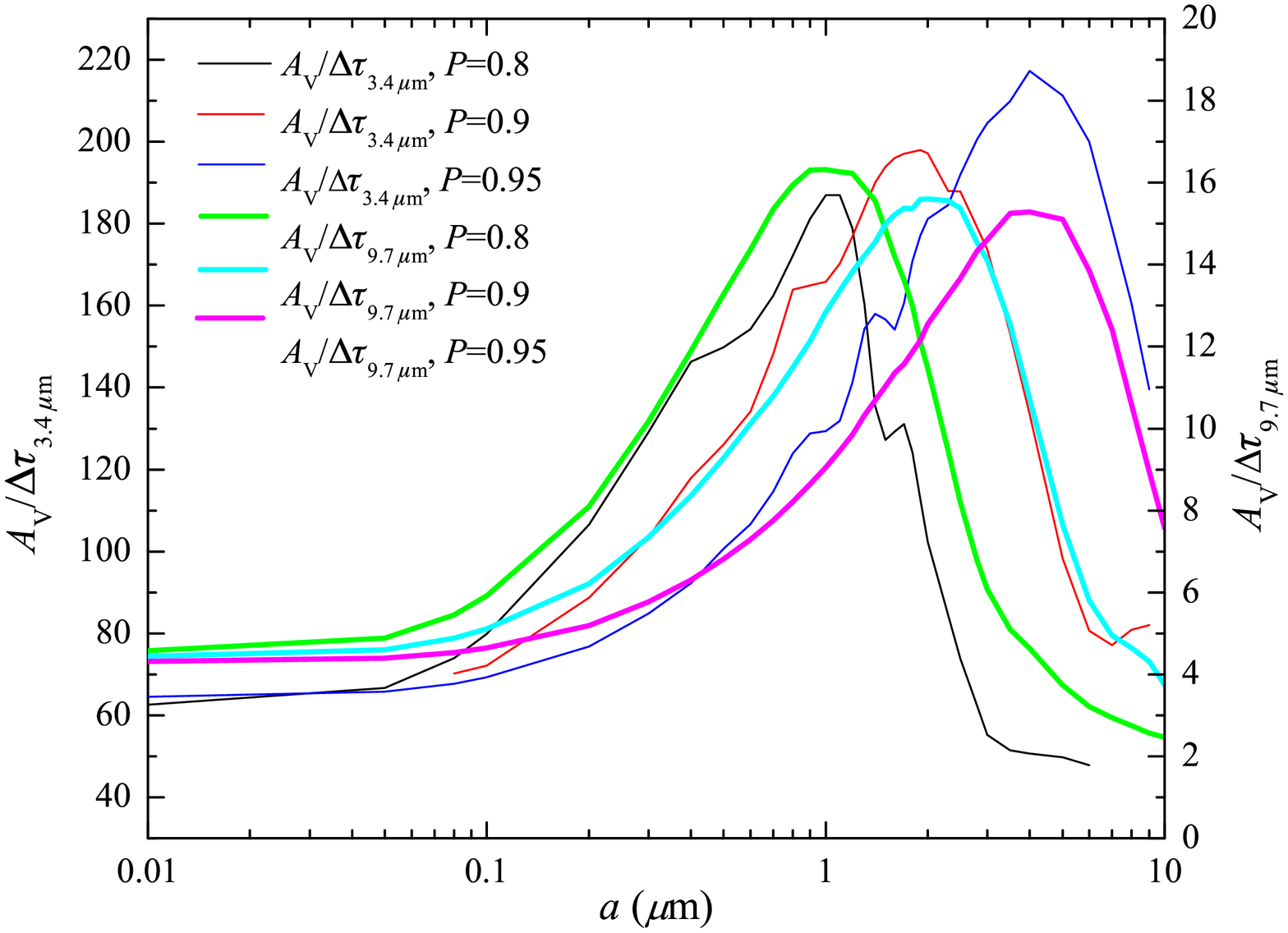}}
\vspace{-1.0cm}
\caption{%
         Variations of $A_V/\tauahc$ and $A_V/\tausil$
         with dust size.
         \label{fig4}
         }
         \vspace{-0.2cm}
\end{figure}
With an increased silicate mass fraction, say, $m_{\rm carb}/m_{\rm
sil} = 0.5$ which is consistent with the {\it in situ} measurements
of comet Halley (Jessberger \& Kissel 1991) and widely adopted in
cometary dust modeling (Greenberg 1998; Greenberg \& Li 1999; Kimura
et al.\ 2006, 2008a; Kolokolova et al.\ 2004, Kolokolova \& Kimura
2008; Mann et al.\ 2006), we obtain $A_V/\tauahc\approx252$ and
$A_V/\tausil\approx27.1$ for the local ISM (assuming compact dust),
and $A_V/\tauahc\approx154$ and $A_V/\tausil\approx11.3$
for the GC (assuming porous dust).
These values are closer to that observed.
It is expected that with a smaller $m_{\rm carb}/m_{\rm sil}$
(i.e. a larger silicate mass fraction), one would obtain a smaller
$A_V/\tausil$ while $A_V/\tauahc$ does not change much. Thus the
observed variations of $A_V/\tauahc$ and $A_V/\tausil$ from the
local ISM to the GC could be explained.
It is worth noting that, based on a detailed analysis of the GC
5--8$\mum$ absorption spectra obtained from the Kuiper Airborne
Observatory, Tielens et al.\ (1996) argued that silicate dust may
contribute as much as 60\% of the interstellar dust volume. This
would translate to $m_{\rm carb}/m_{\rm sil} \approx 0.34$ if we
assume that the remaining 40\% of the interstellar dust volume is
all from the 3.4$\mum$ C--H feature carrier (which is indeed a
very generous assumption).

Admittedly, the proposed explanation is oversimplifed. In the future
we will consider more realistic models in which more dust species
(e.g. hydrogenated amorphous carbon with a range of C/H ratios), the
distribution of dust along the line of sight toward the GC (e.g. see
Sandford et al.\ 1995), a distribution of dust sizes, and the
possible porous nature of the diffuse ISM dust (e.g. see Mathis \&
Whiffen 1989) will be considered.

\acknowledgments{We thank the anonymous referees for their very
helpful comments. We thank Dr. H. Kimura for sharing with us his
work prior to publication. This work is supported in part by NSFC
grant No.\,10603001, SRFDP grant No.\,20060027013 and NCET-05-0144.
AL is supported in part by NASA/HST Cycle-15 Theory Program and NSF
grant AST\,07-07866.}


\email{J.~GAO (e-mail: jiangao@bnu.edu.cn)}, B.~W.~Jiang (e-mail:
bjiang@bnu.edu.cn), and A. Li (e-mail:
lia@missouri.edu)\label{finalpage} \lastpagesettings

\end{document}